\documentclass[preprint,twoside,latterpaper,showpacs,floatfix]{revtex4} 
\usepackage{graphicx}   
\usepackage{epsfig}
\usepackage{latexsym}

\begin{document} 
\title{The CH$_{3}$SH molecule deposited on Cu(111) and deprotonation } 
\author{Jian-Ge Zhou, Quinton L. Williams and Frank Hagelberg}   
\affiliation{Department of Physics, Atmospheric Sciences, and Geoscience, Jackson 
State University, Jackson, MS 39217, USA }  
\author{}  
\affiliation{} 

\begin{abstract}
We demonstrate for the first time that when
a methanethiol adsorbed on the regular Cu(111) surface, the dissociative structure is thermodynamically more 
stable than the intact one. The computational results show that at low temperature the methanethiol adsorbate prefers 
the atop site of the regular Cu(111) surface.
As the temperature is increased, the S-H bond is broken and the methylthiolate favors the hollow sites.  
On the defected Cu(111) surface,  the dissociative configuration is still thermodynamically more 
stable than the nondissociative one. The calculation indicates that the hydrogen 
initially attached to the sulfur would like to form a bond with the copper
surface rather than desorb from it. Even though both copper and gold are the noble metal,
the stability of the methanethiol adsorption
on the Cu(111) substrate is almost the reverse of that on the Au(111).

\end{abstract}  
\maketitle   

\section{Introduction}

In recent years, a lot of work has been conducted on the interaction of alkanethiols 
with gold surface because of their ability to form 
self-assembled monolayers (SAMs); see Ref. \cite{vvb} for a review. These SAM films have wide
application on wetting phenomena, tribology, chemical and biological sensing, 
optics and nanotechnology. The consistent model for the methanethiol adsorption process
on the Au(111) substrate was proposed recently \cite{rlm,zh}.  It was 
revealed  that the methanethiols
stay intact on the regular Au(111) surface, but the S-H bond ruptures
on the defected Au(111) surface. On other noble metals such as
copper, the adsorption pattern of the
alkanethiols (or alkanethilates) on the Cu(111) is distinct from that on the Au(111) 
surface \cite{jwj}-\cite{fpp}. 
Then one may ask how the S-H bond in the methanethiol on 
the regular (or defected) Cu(111) surface breaks. 

The normal incidence X-ray standing wave measurements show that at the low
temperature, the intact methanethiols prefer a local
atop site on the Cu(111) surface, but at 140K methanethiols turn into 
methanethiolates depositing on unreconstructed
areas of the Cu(111) surface, occupying two nonequivalent hollow sites with similar probability \cite{jwj}. 
At room temperature, when methanethiols or dimethyl disulfides adsorbed on the
Cu(111), the formation of the pseudosquare reconstruction 
was observed via scanning tunneling microscopy \cite{dw1,dw2}.
A grazing incidence surface X-ray study indicates that 
intact alkanethiols favor hollow sites on Cu(111) \cite{fsd}. However,
 there has been no experimental proof for hydrogen desorption from such processes.
As we know, alkanethiols only stay intact on the regular Au(111) surface, but the S-H bond ruptures
on the defected Au(111) surface \cite{rlm,zh}. So far, no theoretical work on 
the scission of the methanethiol on the Cu(111) surface has been done,
thus it is imperative to study the mechanism of the S-H bond rupture theoretically. 

In this paper, we study the adsorption of methanethiol 
on the Cu(111) surface by the density functional theory \cite{khf}.
First, we will present adsorption energies and geometries for the nondissociative and dissociative
configuration on the regular Cu(111) surface. We demonstrate, for the first time, 
the dissociative structure is the most stable one for the regular Cu(111) surface, which is opposite to 
the case when the methanethiols adsorbed on the regular Au(111) \cite{rlm,zh}.  
The S atom of the intact methanethiol
stays on the top of the copper atom of the surface. The energy barrier is about 0.2 eV. 
This implies that at low temperature, the intact methanethiol prefers a local
atop site on the regular Cu(111) surface \cite{jwj}. As the temperature increased, 
the energy barrier can be overcome and the hydrogen is detached from the S atom, which is consistent with
the experimental observation \cite{jwj,dw2,fsd}. To discuss the stability of configurations for 
the defected Cu(111) surface, we assume a vacancy in the top
layer of the Cu(111) surface. We find that on the defected surface, the dissociative structure
is still more stable than nondissociative one. The hydrogen initially attached to the sulfur forms
a bond with the Cu(111) substrate instead of desorbing from it.

\section{Computational Method}

The calculations were performed by the VASP code \cite{khf} in two approaches: 1) using the projector
augmented wave potential \cite{kj,peb} and the Perdew-Wang 91 (PW91) generalized gradient approximation to exchange
and correlation (XC) \cite{pw};
2) the ultrasoft pseudopotential \cite{khf}  with XC based on the Perdew-Burke-Ernzerhof (PBE) formalism \cite{pbe}.
The wave functions are expanded in a plane wave basis with an energy cutoff of 400 eV. The Brillouin zone integration 
is performed by use of the Monkhorst-Pack scheme \cite{mp}. 
We utilized a $3\times 3\times 1$  k point mesh for the geometry optimization.
The adopted supercell corresponds to
12 Cu atoms per layer. The Cu atoms in the top three atomic layers are allowed to relax, 
while those in the bottom layer are fixed to simulate bulk-like termination \cite{zwh}. 
The vacuum region comprises seven atomic layers, which
exceeds substantially the extension of the methanethiol molecule.
To examine the accuracy of our approach we increased the energy cutoff to 500 eV and the number of 
k points to $8\times 8\times 1$. In both cases, the difference amounted
to less than $1.8\%$. We further calculated the copper lattice constant, 
and found it to agree with the experimental value \cite{rw} within  $0.6\%$. 

\section{Results and Discussion}
\subsection{Methanethiol molecule deposited on the regular Cu(111) surface}

\begin{figure}
\includegraphics[width=4.9in]{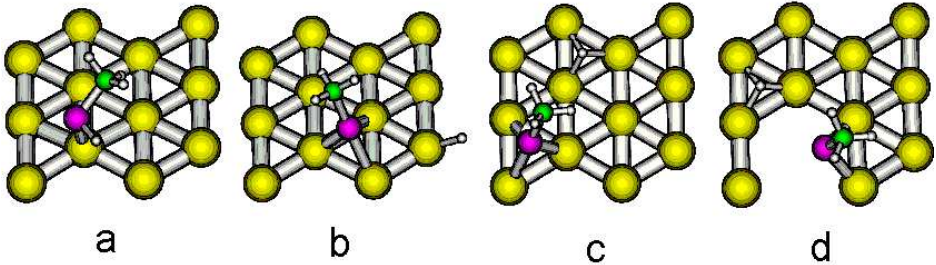}
\caption{(color online). (a) Intact structure on the regular Cu(111). (b) Dissociative structure in the
fcc hollow site of the
regular Cu(111) (fcc-bri). (c) Dissociative structure in the hcp hollow site (hcp-bri).
(d) Dissociative structure on the Cu(111) with a vacancy.} 
\label{structure}
\end{figure}

We begin with the geometries and adsorption
energies of the optimized structures for the methanethiol on the Cu(111) surface at 
0.25 ML, as displayed in Table \ref{geom}. 
\begin{table}
\caption{The geometries and adsorption energies for the optimized nondissociative and
dissociative configurations 
on the regular and defected Cu(111) at 0.25 ML. The entries S site,
$\theta$, tilt, $d_{S-Cu}$ ($\AA$) and $E_{ads}$ (eV) refer to the S headgroup atom
site, the angle between the S-C bond direction and the normal to the Cu(111), the region
of th S-C bond tilted,  
the shortest S-Cu bond length and the adsorption energy, respectively.}
\begin{center}
\begin{tabular}{cccccc}  
\hline
\hline
Configuration&S site &$\theta$&tilt& $d_{S-Cu}$ & $E_{ads}$\\
\hline
Regular Surface&~&~&~&~&~\\
intact&top&65.6 &hcp&2.43&0.61\\
dissociative&fcc-bri&40.2 &hcp&2.26&1.14\\
dissociative&hcp-bri&40.6 &fcc&2.28&1.11\\
methylthiolate&fcc-bri&46.2 &hcp&2.28&2.76\\
$(CH_{3}S)$&~&~&~&~&~\\
Defected Surface&~&~&~&~&~\\
intact&top&54.0 &fcc&2.38&0.74\\
dissociative&two bonds with &16.6 &hcp&2.29&1.64\\
~&two Cu atoms&~ &~&~&~\\
\hline
\end{tabular}
\end{center}
\label{geom}
\end{table}
The symbol fcc-bri (or hcp-bri) in Table \ref{geom} represents that the S atom 
is placed in a fcc (or hcp) hollow region, but leaned toward the bridge.
The stable configurations on the Cu(111) surface are shown
in Fig. \ref{structure}.

We calculated nineteen nondissociative configurations on the regular Cu(111) surface, which
are characterized by the S atom adsorption site (atop, bridge, fcc and hcp
hollow), the angle of the S-C bond, and
the direction of the H-S bond. The 
energy difference among the configurations is around 0.1 eV, i.e., the adsorption energies
span 0.30 eV - 0.61 eV. However, the adsorption energies weakly depend on the tilting of the alkyl chain,
which reveals the ionic nature of the bonding. 
The first row of Table \ref{geom} displays that the adsorption energy for the most stable nondissociative 
structure is 0.61 eV
(Fig. \ref{structure}a), and the S adsorption site is on the top of the Cu atom. 
Thus at low temperature, the intact methanethiol stays on the top site of
 the regular Cu(111) surface, in keeping with a recent experimental 
result \cite{jwj}. The second and third rows of Table \ref{geom} 
list two most stable  dissociative structures on the regular Cu(111) substrate, and
their adsorption energies are 1.14 eV and 1.11 eV.  The 
S atom for both configurations are in hollow site, i.e., fcc-bri (Fig. \ref{structure}b) and
hcp-bri (Fig. \ref{structure}c), which is in favor of 
the explanation in Ref. \cite{krs} (when
dynamical effects are considered, only the hollow site occupancy is observed \cite{krs}). 
The first and second rows in Table \ref{geom} reveal that
the adsorption energy
of the dissociative structure is higher than that of the nondissociative one by 0.53 eV.
To further examine this result, we apply the ultrasoft pseudopotential in conjunction with the 
PBE XC \cite{pbe} to recalculate the above structures, and find an adsorption energy difference of 0.51 eV between
the cases of cleavage and noncleavage. Thus we conclude that the dissociative configuration
is more stable than the nondissociative one, which is consistent with the 
experimental observation that
at low temperature, the intact methanethiol stays on the regular Cu(111) surface, and when the temperature increased,
methanethiolates deposited on unreconstructed
areas of the Cu(111) surface, occupying two nonequivalent hollow sites with similar probability \cite{jwj}. 

\subsection{Projected density of states}

\begin{figure}
\includegraphics[width=5.3in]{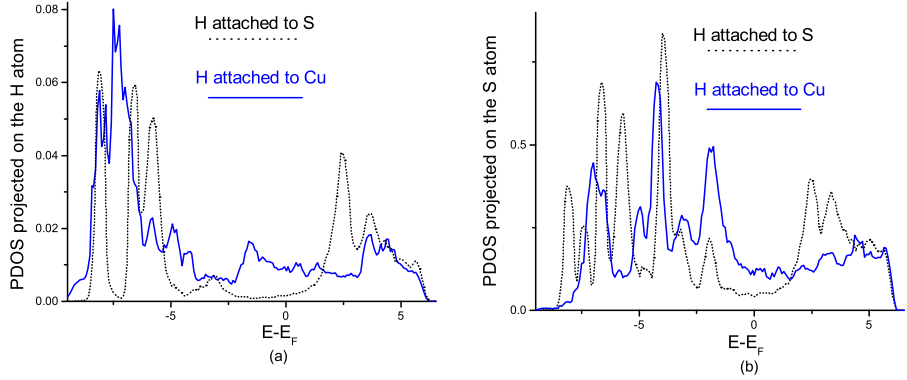}
\caption{(color online). (a) The PDOS projected on the hydrogen atom initially attached to the S atom adsorbed on the
regular Cu(111) surface. (b) The PDOS projected on the S atom adsorbed on the
regular Cu(111) surface.}  
\label{dos}
\end{figure}

To see why the dissociative structure is more stable than the nondissociative one, we calculated the partial
density of states (PDOS) for the H atom initially attached to the sulfur atom (Fig. \ref{dos}a). 
The PDOS projected on the S atom adsorbed on the
regular Cu(111) surface is illustrated in Fig. \ref{dos}b. Fig. \ref{dos}a displays that
the primary peaks of the hydrogen PDOS for the dissociative structure are at positions of lower energy than those
for the nondissociative one. The electronic configuration related to the former case is thus more stable than
that corresponding the latter, and the hydrogen atom forms a stronger bond with the Cu(111) surface than with the
sulfur atom. 

\subsection{Interacting bond}

\begin{figure}
\includegraphics[width=4.2in]{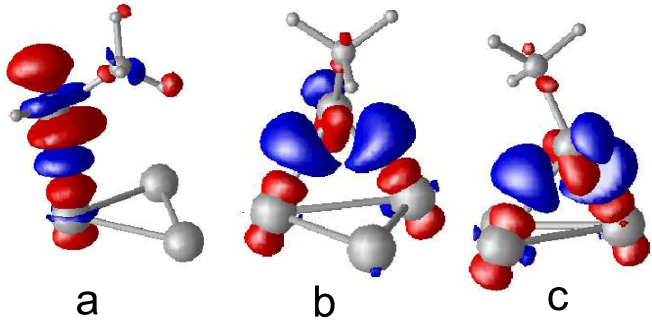}
\caption{(color online). The isosurfaces of the charge-density difference for
(a) nondissociative adsorption with blue (accumulation of electrons)/red (depletion 
of electrons) isosurface value: $\pm$0.015 e/${\AA}^3$, 
(b) dissociative structure (front view) and (c) dissociative structure
(back view) with isosurface value: $\pm$0.025 e/${\AA}^3$. Only three related copper atoms 
of the Cu(111) surface are illustrated.}
\label{chg}
\end{figure}

To elucidate the interacting bond between the
methanethiol (or methylthiolate) and the Cu(111) substrate,
we evaluate the spatially resolved charge-density difference:
$\Delta\rho(\vec{r}) = \rho_{ads/sub}(\vec{r})-\rho_{sub}(\vec{r})-
 \rho_{ads}(\vec{r})$, where $\rho_{ads/sub}$, $\rho_{sub}$ and 
 $\rho_{ads}$ are the electron charge densities of the relaxed adsorbate-substrate
system, of the clean relaxed surface, and of the isolated but
adsorption like deformed adsorbate (without substrate), respectively. The isodensity surfaces of the charge-density
difference for the intact and dissociative structure 
are depicted in Fig. \ref{chg}a, Fig. \ref{chg}b and Fig. \ref{chg}c (to discuss the change of the S-Cu bond before
and after dissociation, in Fig. \ref{chg}, 
we display only the surrounding part of the S-Cu bond).

\begin{figure}
\includegraphics[width=5.3in]{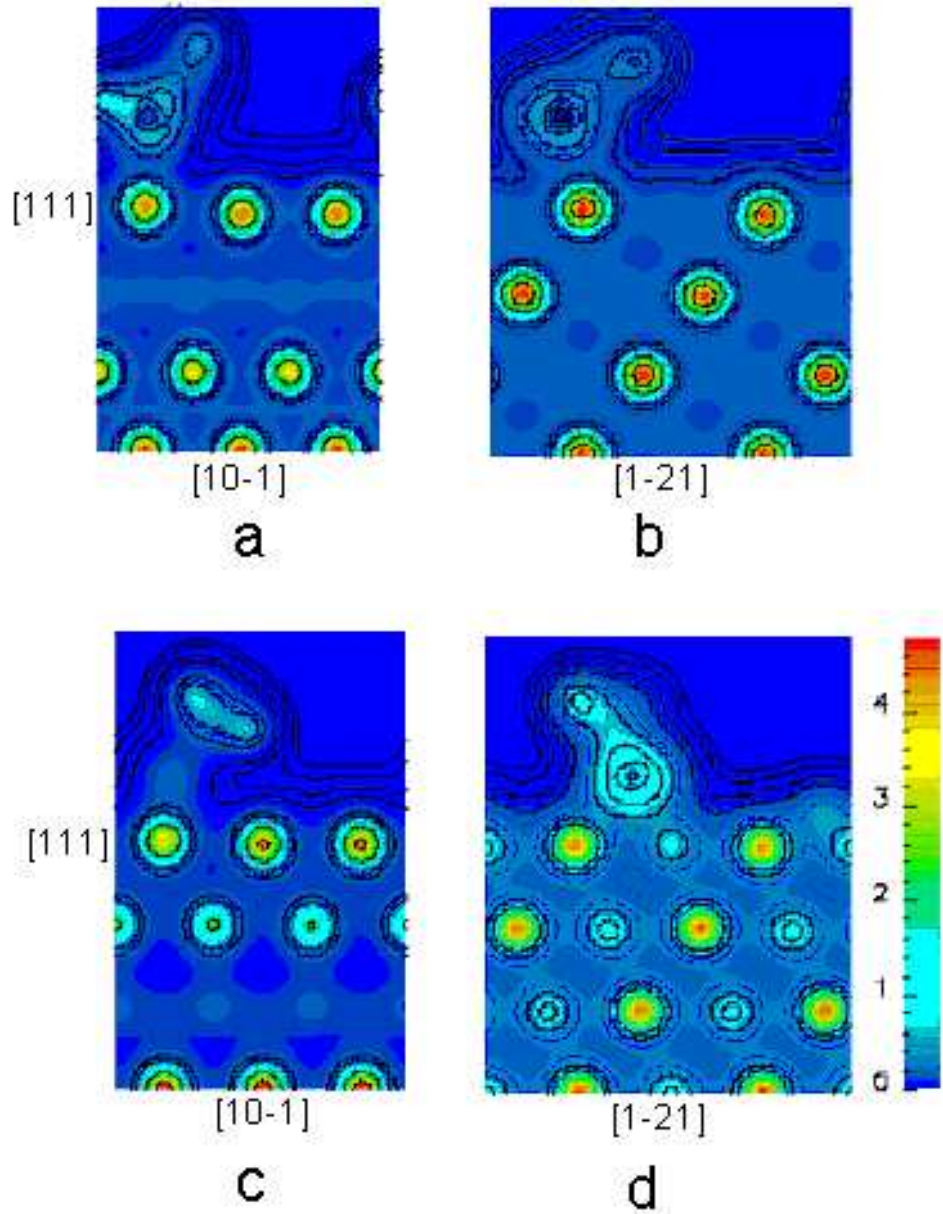}
\caption{(color online). The contour plots of the charge densities.
(a) Charge density plot in (1-21) plane which extends along [111] and [10-1]
directions and (b) in (10-1) plane for the intact adsorption.
(c) Charge density plot in (1-21) plane and (d) in (10-1) plane for the dissociative adsorption.}
\label{contour}
\end{figure}

The calculated charge transfers from the Cu(111) substrate
to the methanethiol for the nondissociative adsorption
and to the methylthiolate for the dissociative structure are 
0.3e and 0.6e, respectively. Fig. \ref{chg}a shows
that for the intact adsorption, the electrostatic
interaction responsible for the bonding  comes from the monopole term (the
0.3e charge transfer) and the dipole moments in the adsorbate and substrate \cite{psb}.
Around the S atom, there is a ``ring" of accumulation of electron charge.
The electrostatic interaction is dominated by the
attractive ionic term modified by a repulsive dipolar term \cite{psb}. The sulfur 
stays on the top of the copper atom, i.e., the sulfur only forms a bond with
one Cu atom. 
The interaction between the methylthiolate and Cu(111) is the ionic
bonding \cite{fpm,fpp}. The sulfur in the methylthiolate forms bonds with 
two copper atoms of the Cu(111) surface (see Fig. \ref{chg}b and Fig. \ref{chg}c),
and its charge transfer (0.6e) is two times of that for the nondissociative case (0.3e). 
It indicates that the methylthiolate
takes in 0.3e electron charge from each copper atom channel (0.3e $\times$ 2 = 0.6e), which
matches with the fact that the adsorption energy
for the dissociative structure (1.14eV) is about two times of that for intact adsorption
(0.61eV).
The charge densities in the [1-21] and [10-1] plane for intact and 
dissociative adsorption are shown in Fig. \ref{contour}, which is consistent with the picture obtained from
the isosurfaces of the charge-density difference, charge transfers, and bond structures.

\subsection{Methanethiol molecule deposited on the defected Cu(111) surface}

On the regular Cu(111) surface, we found that the dissociative structure
is thermodynamically more stable than the nondissociative one. To see if the defects can change the above
conclusion, we consider the Cu(111) surface with a vacancy in the top layer of the supercell. 
The geometries and adsorption energies for the optimized nondissociative and dissociative
configurations on the Cu(111) surface with a vacancy are indicated in the fifth and sixth rows of 
Table \ref{geom}. It shows that the adsorption energy for the
dissociative structure (Fig. \ref{structure}d) is 0.9 eV higher than that for the intact one, i.e., the 
dissociative configuration on the defected Cu(111) surface is thermodynamically more 
stable than the nondissociative one. For the stable dissociative adsorption,
the S atom forms two bonds with two adjacent copper atoms which surround the vacancy.

\subsection{Energy barrier}

\begin{figure}
\includegraphics[width=6.2in]{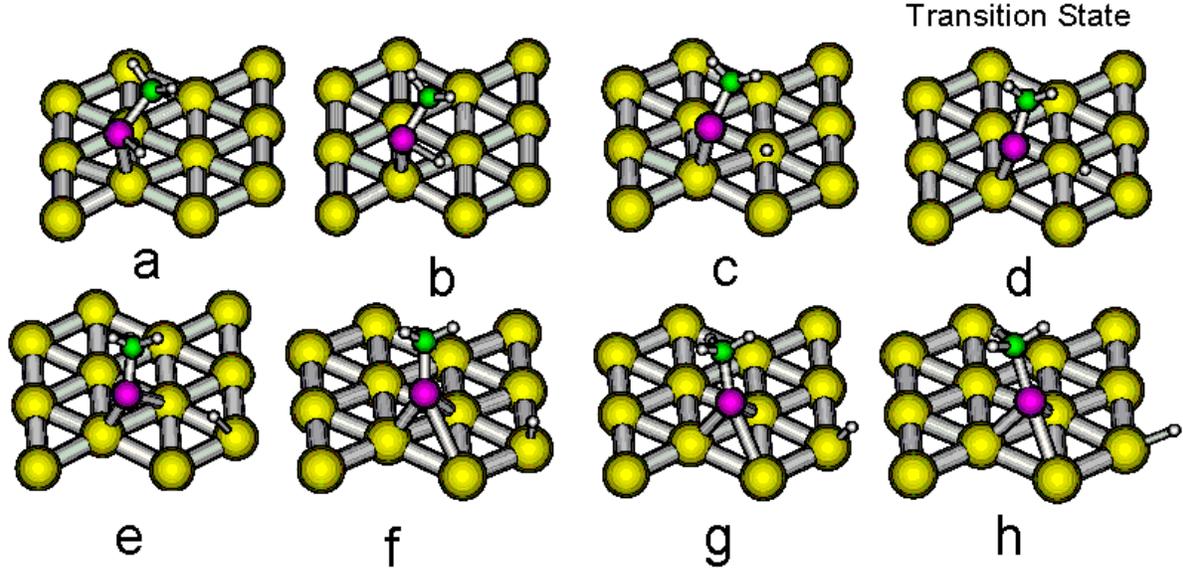}
\caption{(color online). The structures for the initial intact adsorption, the final dissociated one, and
the six images.}  
\label{image}
\end{figure}

To explain why at low temperature, the intact methanethiol on
the regular surface was observed, and around 140K, the S-H bond in
the methanethiol on the Cu(111) was broken, we calculated the energy
barrier for the process
that lead from the nondissociative structure Fig. \ref{structure}a to the
dissociative structure Fig. \ref{structure}b. The energy barrier was calculated by 
the climbing nudged elastic band method, \cite{huj}-\cite{jmj}
where six equidistant images have been used. The atomic configurations for the
initial nondissociative adsorption, 
the six intermediate images, and the final dissociative one are shown in Fig. \ref{image}. The structure of the
transition state is depicted in Fig. \ref{image}d.  

\begin{figure}
\includegraphics[width=3.1in]{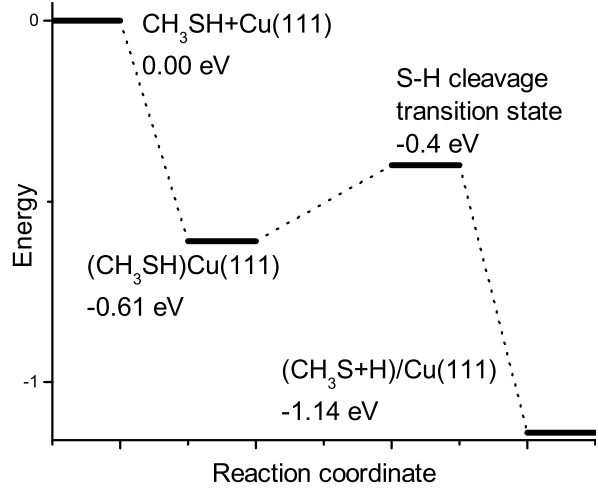}
\caption{The relative energy levels along the S-H bond scission reaction paths.}  
\label{energy}
\end{figure}

The relative energies along the S-H bond
cleavage reaction path are schematically illustrated in Fig. \ref{energy}. The 
energy barrier is about 0.2 eV, which is lower than the
corresponding energy barrier for the scission of the S-H bond in a single methanethiol molecule. 
The reason is that when the methanethiol is adsorbed on the Cu(111), the surface plays a role as a catalyst, which can
suppress the barrier height. As the temperature is raised, the S-H bond break reaction is activated
over the energy barrier, which is consistent with the experimental 
measurement \cite{jwj}. 

\subsection{Hydrogen atom behavior}

We have seen that at low temperature, the intact methanethiol favors the atop site of the 
Cu(111) surface, and when the temperature increased, the S-H bond in the methanethiol on the Cu(111) surface
is ruptured. The hydrogen
could either stick to the Cu(111) surface or desorb in molecular form. No experimental
evidence of either process has been found. To predict if the hydrogen forms a bond
with the Cu(111) surface or desorbs in the hydrogen molecule, we compared the 
interaction energy for the dissociative structure $E_{(CH_{3}S+H)/Cu(111)}$ with that
for methylthiolate on the Cu(111) and the desorbed hydrogen molecule, i.e., $E_{(CH_{3}S)/Cu(111)}$ +
$\frac{1}{2}E_{H_2}$. We found that $E_{(CH_{3}S+H)/Cu(111)}$ is lower than $E_{(CH_{3}S)/Cu(111)}$ +
$\frac{1}{2}E_{H_2}$ by 0.24 eV, which implies that
the hydrogen would like to form a bond with copper
surface rather than desorb from the surface.

To check if the deposited hydrogen modified the methylthiolate adsorption pattern, we calculated 
fifteen methylthiolate structures on the regular Cu(111) surface. The most stable structure
is shown in the fourth row of Table \ref{geom}. The second and fourth rows in the Table  
display that the S-Cu bond lengths and the angles $\theta$
for the dissociative structure (both the hydrogen and the CH$_3$S  adsorbed on the Cu(111) surface) and for
the methylthiolate (only CH$_3$S) on the Cu(111) are 2.26 $\AA$, 40.2$^{\circ}$, 2.28 $\AA$ and 46.2$^{\circ}$,
respectively.  
The hydrogen adsorbed to Cu(111) modifies the methylthiolate adsorption structure by 6$^{\circ}$.

\subsection{Comparison with the methanethiol adsorption on the Au(111) surface}

\begin{figure}
\includegraphics[width=3.9in]{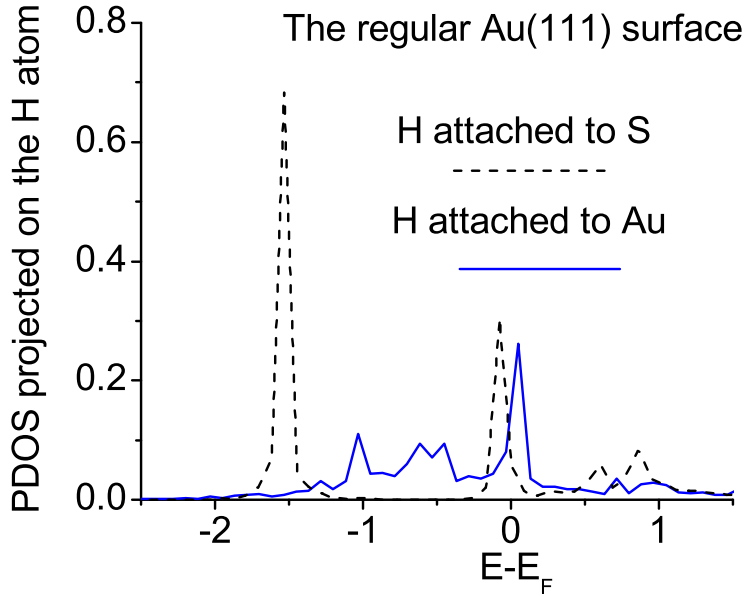}
\caption{(color online). The PDOS projected on the hydrogen atom initially attached to the S atom 
adsorbed on the regular Au(111) surface.}  
\label{audos}
\end{figure}

We compare the methanethiol adsorption process on the Cu(111) substrate with that on the other noble 
metal surface - Au(111). On the regular Cu(111) surface, the methanethiol dissociative adsorption is
more stable than the intact one. The corresponding energy barrier is about 0.2 eV. 
So at low temperature, the methanethiol stays intact on the atop site of the regular Cu(111). When
the temperature is raised, this energy barrier is overcome, and the S-H bond breaks. The corresponding
hydrogen is adsorbed on the surface. The presence of the defects on the Cu(111) does not change the situation.
However, on the regular Au(111) surface, the nondissociative structure is more stable than 
the dissociative one. At low temperature, the methanethiol favors the atop site of the regular Au(111).
As the temperature is increased, the methanethiol molecule desorbs from the regular Au(111). 
For the defected Au(111), this order of stabilities is reversed. 
To unveil the origin of the difference between the adsorption on the Cu 
and Au surface, we calculated the partial
density of states (PDOS) for the H atom initially attached to the sulfur atom
adsorbed on the regular Au(111) surface [see Fig. \ref{audos}]. 

Comparing Fig. \ref{audos} with Fig. \ref{dos}a, we have found that for 
the Cu surface the primary peaks of the hydrogen 
PDOS for the dissociative structure are at positions of lower energy than those
for the nondissociative one. However, for the Au surface, 
the primary peaks of the hydrogen PDOS for the intact structure are at positions of lower energy than those
for the dissociative one. Thus such a comparison explained the physical reason of the difference between 
the Cu and Au surface. 

\section{Summary}

In conclusion, we have demonstrated for the first time that when
a methanethiol adsorbed on the regular Cu(111) surface, the dissociative structure is thermodynamically more 
stable than the intact one. The energy barrier 
for the process that lead from the nondissociative structure to the
dissociative one is about 0.2 eV. Thus at low temperature, the methanethiol adsorbate prefers 
the atop site of the regular Cu(111) surface \cite{jwj}.
As the temperature rises, the S-H bond is broken, and the methylthiolate favors the hollow sites \cite{jwj}.  
On the defected Cu(111) surface,  the dissociative configuration is still thermodynamically more 
stable than the nondissociative one. The calculation shows that the hydrogen 
initially attached to the sulfur would like to form a bond with copper
surface rather than desorb from it. Even though both copper and gold are the noble metal,
the stability of the methanethiol adsorption
on the Cu(111) substrate is almost the reverse of that on the Au(111).

\section*{Acknowledgments}

We thank Dr. C. Xiao for a helpful discussion. 
This work is supported by the NSF through Grants HRD-9805465 and DMR-0304036, 
NIH through Grant S06-GM008047, and DoD through Contract $\#$W912HZ-06-C-005.


\begin{thebibliography}{99}   

\bibitem{vvb} C. Vericat, M. Vela, G. Benitez, J. Martin Gago, X. Torrelles,
R. Salvarezza, J. Phys.: Condens. Matter 18, R867 (2006).
\bibitem{rlm} I. Rzeznicka, J. Lee, P. Maksymovych, J. Yates, Jr., 
J. Phys. Chem. B109, 15992 (2005).
\bibitem{zh}  J. Zhou, F. Hagelberg, Phys. Rev. Lett. 97, 045505 (2006).
\bibitem{jwj}  G. Jackson, D. Woodruff, R. Jones, N. Singh, A. Chan, B. Cowie, V. Formoso,  
Phys. Rev. Lett. 84, 119 (2000).
\bibitem{dw1} S. Driver, D. Woodruff, Surf. Sci. 457, 11 (2000).
\bibitem{dw2} S. Driver, D. Woodruff, Langmuir 16, 6693 (2000).
\bibitem{fsd} P. Floriano, O. Schlieben, E. Doomes, I. Klein, J. Janssen,
J. Hormes, E. Poliakoff, R. McCarley, Chem. Phys. Lett. 321, 175 (2000).
\bibitem{ank} Y. Akinaga, T. Nakajima, K. Hirao, J. Chem. Phys. 114, 8555 (2001).
\bibitem{tpk} R. Toomes, M. Polcik, M. Kittel, J. Hoeft, D. Sayago,
M. Pascal, C. Lamont, J. Robinson, D. Woodruff, Surf. Sci. 513, 437 (2002).
\bibitem{fpm}  A. Ferral, P. Paredes-Olivera, V. Macagno, E. Patrito, 
Surf. Sci. 525, 85 (2003).
\bibitem{okk} M. Ohara, Y. Kim, M. Kawai, Langmuir 21, 4779 (2005).
\bibitem{krs}  M. Konopka, R. Rousseau, I. Stich, D. Marx,  
Phys. Rev. Lett. 95, 096102 (2005).
\bibitem{cpm} F. Cometto, P. Paredes-Olivera, A. Macaguo, E. Patrito,
J. Phys. Chem. B109, 21737 (2005).
\bibitem{pmq}  G. Parkinson, M. Munoz-Marquez, P. Quinn, M. Gladys,
D. Woodruff, P. Bailey, T. Noakes,  Surf. Sci. 598, 206 (2005).
\bibitem{fpp}  A. Ferral, E. Patrito, P. Paredes-Olivera, J. Phys. Chem. B110, 17050 (2006).
\bibitem{khf}   G. Kresse,  J. Hafner,  Phys. Rev. B47, R558 (1993); G. Kresse and J. Furtmuller, 
Phys. Rev. B54, 11169 (1996).
\bibitem{kj}   G. Kresse, J. Joubert,  Phys. Rev. B59, 1758 (1999). 
 \bibitem{peb}   P.E. Blochl, Phys. Rev. B50, 17953 (1994). 
 \bibitem{pw}   J. Perdew, Y. Wang, Phys. Rev. B46, 6671 (1992).
\bibitem{pbe}   J. Perdew, K. Burke, M. Ernzerhof, Phys. Rev. Lett. 77, 3865 (1996).
 \bibitem{mp}   H. J. Monkhorst, J. D. Pack, Phys. Rev. B13,5188 (1976).
\bibitem{zwh}  J. Zhou, F. Hagelberg, C. Xiao, Phys. Rev. B73, 155307 (2006); 
J. Zhou, Q. Williams, F. Hagelberg, Phys. Rev. B76, 075408 (2007).  
\bibitem{rw}  R. Wyckoff, Crystal Structure, 2nd ed. (Interscience, New York, 1963).
\bibitem{psb}  M. Preuss, W. Schmidt, F. Bechstedt, Phys. Rev. Lett. 94, 236102 (2005).
\bibitem{huj} G. Henkelman, B. Uberuaga, H. Jonsson, J. Chem. Phys. 113, 9901 (2000).
\bibitem{mjs} G. Mills, H. Jonsson, G. Schenter, Surf. Sci. 324, 305 (1995).
\bibitem{jmj} H. Jonsson, G. Mills, K. Jacobsen, in Classical and
Quantum Dynamics in Condensed Phase Simulations, edited by B. Berne, G. Ciccotti,
D. Coker (World Scientific, Singapore, 1998).


  




\end{thebibliography}
\end{document}